\documentclass{aastex}
\usepackage{spr-astr-addons}
\usepackage{url}

\RequirePackage{color}

\begin{document}

\title{Identifying Two Newly Discovered Black Hole X-ray
Transients in the Galactic Center}

\author{Byeong Yeon Ryu}

%\altaffiltext{1}{The Hotchkiss School}
%\altaffiltext{2}{Princeton High School}
%\altaffiltext{3}{Columbia University in the City of New York}

\begin{abstract}
We analyze the two unclassified newly outbursting X-ray binary
transients in the galactic center, \emph{Swift J174540.7-290015} and
\emph{Swift J174540.2-290037}, in an attempt to identify the nature of
these sources. We engage in a thorough spectral analysis for the two
systems, by examining the fits of various astrophysical models, in
addition to assessing their light curves and power spectra. Our results
showed that Swift 15 has a low temperature blackbody spectrum with a
hard power-law tail, and Swift 37 has a broad iron line and a small
equivalent width; both transients exhibit classic spectral signatures of
black hole systems. Additionally, our observations for both sources
yielded the absence of type 1 X-ray bursting and the possible presence
of low-frequency quasi-periodic oscillations. Our finding that these two
sources appear to be black hole X-ray systems is a groundbreaking
discovery, as it adds on to an extremely small population of
well-established black hole candidates in the entire galaxy, in addition
to giving the first strong indication for a clustering of black holes
within the central few parsecs of the galactic center.
\end{abstract}

\section{Introduction}

One of the most intriguing challenges in high-energy galactic center
astrophysics, has always been to identify more black hole systems, in
order to account for their curiously small population, contrary to
popular theoretical implications. It is widely believed that as our
galaxy began to progress many billions of years ago, the black holes
should gradually fall closer into the center of the galaxy throughout
time due to their loss of energy as they collide with astrophysical
matter in the interstellar medium (Miralda-Escudé \& Gould 2000). Thus,
theoretical evidence implies millions of black holes in the galaxy;
however, the culmination of all historical observations only yields less
than 70 strong black hole candidates in the entirety of the Milky Way,
with only 21 dynamically confirmed black holes (Tetarenko, 2016). Of
these, only one has ever been identified in the central two to three
parsecs of the center of the galaxy (Tetarenko, 2016). The ongoing
studies of galactic center sources is extremely exciting to this date,
as they add insight to the everlasting uncertainty regarding the
scarcity of identified black hole systems in the galactic center.

X-ray telescope \emph{Swift}, since 2006, has been monitoring the
galactic center on a daily basis, and within these past seven months,
they found very exciting observations of three completely new X-ray
binary transient sources (``The Swift X-ray Monitoring Campaign,'' n.d).
In early February of 2016, they reported the transient source,
\emph{SWIFT J174540.7-290015} (Swift 15\emph{)}, located just under 17
arcseconds north of supermassive black hole Sagittarius A* (positioned
at the center of the galaxy). In late May, another X-ray binary
transient system was detected, \emph{SWIFT J174540.2-290037} (Swift
37)\emph{,} approximately 10 arcseconds south of the galactic center
(``The Swift X-ray Monitoring Campaign,'' n.d). A third X-ray source
detected in early May, thought to have started experiencing accretion
bursts since late March, \emph{GRS 1741-2853,} has already been
identified unambiguously as a neutron-star binary (Rutledge, 2016). Our
research will therefore focus on the identification of Swift 15 and
Swift 37, two fresh and unclassified X-ray transients that are
incredibly close to the galactic center.

Based on the preliminary observations from \emph{Swift}, \emph{NuSTAR}
also became motivated to monitor the region of the galactic center in
which the transients popped up and recorded their respective data. The
\emph{NuSTAR} telescope is internationally renowned as one of the
premier X-ray telescopes due to its large effective energy band, ranging
from 3 keV to 79 keV (``NuSTAR Bringing the High,'' n.d). Because of
this advantage, we will use \emph{NuSTAR}'s observations, as they are
most compatible with high energy imaging software and will yield the
most information-rich spectral results (``NuSTAR Bringing the High,''
n.d).

X-ray binary systems, such as the transients that we will be delving
further into, are the fundamental focus of this subcategory of
astrophysics. The galactic center is populated with a huge variety of
different objects, and telescopes such as \emph{Swift}, \emph{NuSTAR,
Chandra}, \emph{XMM}, and others are able to detect many of these
sources due to the X-ray emission from the source's binary system
(``Astronomer's Toolbox,'' n.d). X-ray binaries are two-star systems
with a compact object and a main sequence companion star. In particular,
X-ray transient binaries have a compact object, typically either neutron
stars or, less commonly, black holes. In these binary systems, the
compact object, with significantly larger gravitational force, will
accrete matter from the companion, or donor star, forming an accretion
disk as the matter gradually falls into the compact object
(``Astronomer's Toolbox,'' n.d). In black hole and neutron star binary
systems, X-rays are detected from the plasma generated from the
differential rotational friction, which occurs due to the difference in
rotation speeds within the accretion disk, alongside the impact of the
accreted matter on the compact object's surface if it is a neutron star;
black holes do not have definitive surfaces (``Astronomer's Toolbox,''
n.d). In this study, we analyze a pair of X-ray binary transients, Swift
15 and Swift 37, that have been dominantly illuminating the field of
view for all X-ray telescopes monitoring the galactic center since their
outbursting six to seven months ago (Rutledge, 2016).

Through a thorough analysis of their spectroscopy, we establish two new
solid black hole transient candidates, both less than one parsec away
from the center of the galaxy. These highly suggestive black-hole
transient systems are the closest black holes to the galactic center
ever recorded. Our results present detailed characteristics of two new
black hole candidates, and we hope our findings on \emph{Swift
J174540.7-290015} and \emph{Swift J174540.2-290037} provide fascinating
insight to the galaxy's black hole population. ~

\section{Methods}

\subsection{\emph{Swift} and \emph{NuSTAR} X-ray Telescope Details}
From the various X-ray telescopes monitoring the galactic center, the
focus of our research will include the initial \emph{Swift}; however,
the main dataset will be provided by the \emph{NuSTAR} images. Although
\emph{Swift} has a relatively lower energy range, from 0.2 keV to 10
keV, it is one of the leading telescopes primarily designed to analyze
the spectroscopy of the most luminous sources in our galaxy, such as
Gamma-ray bursters and X-ray transients as well as time variable sources
(Degenaar, Miller, 2015). \emph{Swift} is able to cover the fluctuating
variability and incoming count-rate range from these extremely bright
sources due to its unique X-ray telescope (XRT). This XRT receives X-ray
emission into a charge-coupled device using 12 multilayer mirrors of a
fairly low-density material, silicon carbide, to utilize the emission
flux for detection of the afterglow (Degenaar, Miller, 2015). X-ray
bursting afterglows have energy ranges precisely that of \emph{Swift}'s
capability. These mechanics enable \emph{Swift} to position sources
accurately with only up to 1 arcsecond radius of error.

\emph{NuSTAR}, on the other hand, has the specialty of having
alternating coatings of high-density materials such as tungsten and
platinum within its numerous layers of mirror coating, used to reflect
incoming X-rays, and low-density materials such as silicon and carbon
(``NuSTAR Bringing the High,'' n.d). These materials, which have high
resistivity, are able to give \emph{NuSTAR} good quantum efficiencies at
high energies, allowing an unmatched maximum range of 79 keV. There are
two detectors on the \emph{NuSTAR} telescope, denoted module A and
module B, that give two sets of data images for each observation
(``NuSTAR Bringing the High,'' n.d).

\subsection{DS9 Extraction and Xspec Modeling}

We first process the \emph{NuSTAR} observations into DS9, a software
used to manipulate the images, or photon count maps, generated by the
telescope detectors (Forster, Grefenstette, Madsen, 2014). With the DS9
images, we create a region of the central 20 arcseconds of the source
and define that as our source region. We put a 20 arcsecond limit, as it
captures around 40\% of the source's photons; however, anything above
this radius might include other contaminating photons. Taking this into
consideration, we also create a background region; an isolated source
spectrum that we will then subtract from the source region photons. The
purpose of this background region is to avoid contamination from other
sources, detected light that does not belong to the source's spectrum
(stray light), and any other diffuse atmospheric emission from the
interstellar medium (Forster, Grefenstette, Madsen, 2014).

Swift 15 and 37, being detected in different locations and times, we
must choose different background regions for them. For Swift 15, we use
the source photons from nearby neutron star ~\emph{AX J1745-2901} source
region as our background file because the transient is located within
supernova remnant, Sagittarius A* East, and we have to account for its
contamination (Rutledge, 2016). For Swift 37, detected later on in late
May, we must account for the source contamination from Swift 15, due to
the closeness of the two transients (Rutledge, 2016). Therefore, we use
the source photons from Swift 15 as our background file for Swift 37.

X-Ray Spectrum, or Xspec, is a widely used astrophysical modeling
language that we will be using to analyze the spectroscopy of these two
transients. Xspec is used so frequently because of its compatibility
with modeling analysis in addition to its convenience on generating
model datasets from \emph{NuSTAR} source and background files (Arnaud,
Keith, Dorman, 2015).

\subsection{Initial Modeling}

With the source and background regions properly defined for Swift 15 and
37, we produced fits with common astrophysical models. The models we
used, given by NASA's astrophysics software (HEASARC), are components
such as \emph{diskbb}, \emph{bbody}, \emph{compbb, powerlaw,} and
\emph{gaussian}. The \emph{diskbb} model calculates the temperature of
the accreting matter in a disk shape, falling into the compact object,
with respect to the radial distance from the compact object, and fits it
to a blackbody spectrum (Arnaud, Dorman, 2015). \emph{Bbody,} a simple a
standard blackbody spectrum model, is an idealized spectrum obtained
from a hot radiator that absorbs all electromagnetic radiation (Arnaud,
Dorman, 2015). The \emph{compbb} model is a blackbody spectrum, modified
with Compton scattering. As jets of X-ray energy are emitted from the
compact object or when a halo of electrons disperses the emitted photons
the inner edge of the accretion disk, this energy collides with the
extremely hot particles within the compact object's corona and scatter
away, forming a \emph{compbb} spectrum (Arnaud, Dorman, 2015). The
\emph{powerlaw} model fits with many X-ray binary systems because as
hard photons are scattered through the compact object's corona and are
reflected into our line of sight from our telescopes, we tend to see a
power law fit (``NuSTAR Bringing the High,'' n.d). Gaussian lines are
used to fit iron-emission lines and possibly other interesting features
within the spectra.

Each spectrum must be modified with the staple multiplicative models of
\emph{tbabs} and \emph{constant}. ~\emph{Tbabs} accounts for the X-ray
absorption from hydrogen molecules in the interstellar medium between
our telescopes on Earth and the sources we are observing (Arnaud,
Dorman, 2015). The \emph{constant} component is an energy independent
factor that accounts for the difference between the effective areas of
module A and module B (Arnaud, Dorman, 2015).

With these models, we will find the best fits for Swift 15 and 37 in
order to determine their properties and to analyze their parameters.
While fitting, we look for the reduced chi-square
(rX\textsuperscript{2}) statistic to be as close as possible to 1, to
determine whether the modeling is accurate. The rX\textsuperscript{2} is
the summation of the residuals of observed and expected model
calculations for the dataset.

Additionally, analyzing the visual spectra of the models is extremely
important, as the residuals and spectral features are vital for our
understanding of the model. We used the \emph{Xwindows} device for the
plots, because it is the premier software for model building on Xspec.

After finding the best fits for the two transients, we use the models to
calculate the equivalent widths, which is the integrated areal photon
flux of the iron emission lines that we expected to see. This result is
extremely important, as different source populations tend to have
accepted general ranges of equivalent widths (Parker, 2016). This can be
done with Xspec using the ``eqwidth'' command.

\subsection{Timing Analysis}

When attempting to find the nature of new X-ray binary systems,
examining the light curves is essential, as its features could be
indicative of the nature of a source population. We created these plots
with count rate as a function of time (s) and used the photons generated
by extracting our source regions from their respective event files. We
used 512 equal newbins, to obtain reasonable bin sizes and to include
the entire timespan of the inputted data. As with the Xspec modeling, we
also used the Xwindows device for plotting.

We did this analysis to see if the transient datasets yield interesting
features in their light curves, namely Type 1 X-ray bursts. This
outbursting occurs periodically when hydrogen and helium burn unstably
on the surface of the compact object; during the duration of the bursts,
the count rate of photons can be detected as extreme outlier peaks
(Galloway, 2008). This will be imperative to analyze, as these bursts
are a definitive characteristic of neutron star systems (Galloway,
2008).

\subsection{Power Spectrum}
Power spectra analysis is one of the most fundamental aspects of looking
into a new source, being a plot of power (erg/s) as a function of source
frequency (Hz). Spectral features of power spectra are studied
intensively by a select group of world-renowned galactic center
astrophysicists, who have produced strong evidence and theories
pertaining to black hole and neutron star classification. In our
analysis of Swift 15 and 37, we looked for the classic black hole X-ray
binary spectral signature of low frequency quasi-periodic oscillations
(QPOs) \textbf{(}Wijnands, van der Klis, 1999). QPOs are an indication
of X-ray emission from the inner edge of the accretion disk, and
detection of this at low frequency, at between 0.01Hz and 1 Hz, is a
widely believed confirmed characteristic of black holes (Wijnands, van
der Klis, 1999).

\subsection{More Sophisticated Reflection Models}
Some xspec models, such as the power law, are used simply to fit the
higher-energy photons, without any astrophysical meaning or relation to
the source population. The more complicated reflection model,
\emph{reflionx}, will be used to take into account the reflection of
high energy photons bouncing off the surface of the accretion disk, or
reabsorbed and emitted energy from the disk (Arnaud, Dorman, 2015). This
pure astrophysical model calculates for this radiation from a power-law
spectrum (Arnaud, Dorman, 2015).

\section{Results}

\subsubsection{Best Fits for Transient \emph{Swift J174540.7-290015}}

\begin{figure}[htpb]
\centering
\includegraphics[width=1\columnwidth]{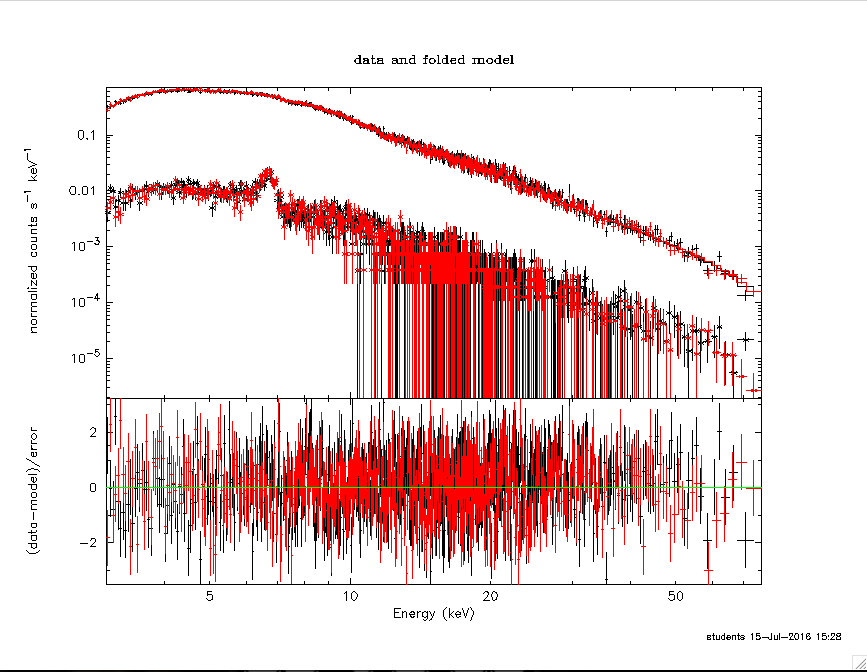}
\caption{Folded spectrum for Swift 15 with the best fit model, with
counts as a function of energy (keV). The datasets, upper emission
(source) and lower emission (background) are in the top region. The
bottom box is the residuals between the model and the data.}
\label{fig_sim}
\end{figure}

As displayed in Fig. 1, we have generated a best fit to model
the data of Swift 15: \emph{tbabs * constant * (diskbb + powerlaw + gaussian)}.
The reduced chi-square statistic of this model is 0.999; the closeness
of this rX\textsuperscript{2} value to 1 is stunning. Looking into its
parameters, we see very interesting results. The power-law photon index
gives a relatively hard value of 1.888, and the \emph{diskbb} component
yields a physically reasonable temperature of 0.967 keV. Many
theoretical and observational astrophysical perspectives have deduced
that between approximately 1.8 to 2.0 in photon index, a transient
source is currently transitioning between a hard and soft state; thus,
we see a presence of a \emph{diskbb} component, here (Tetarenko, 2016).

Comparing with the results of known neutron star binaries, which
typically have disk temperatures around 1.5 keV and higher, we find
Swift 15 yields a significantly lower blackbody temperature. Having a
low disk blackbody characteristic temperature in addition to a hard
power-law tail is a strong indication of the X-ray spectral signature of
a black hole (Tetarenko, 2016).

The gaussian fitting produced, as expected, a neutral iron line at 6.444
keV with a relatively broad line width (Sigma) of 1.904 keV. Equivalent
width calculations yielded 1.461 keV.

Looking into the spectrum, we see very reasonably oscillating residuals,
with no sharp peaks or dips. With the smooth residuals between 6 and 7
keV, we notice the gaussian almost perfectly accounting for the iron
emission. At around 8 -10 keV and above 20 keV, we do not see any
spectral signatures of cyclotron lines; these features are classic
indication of neutron stars (Tetarenko, 2016).

With the tremendous fit with a disk blackbody and power-law
characteristics, we are motivated to fit with a \emph{reflionx} model,
for a better astrophysical understanding of the the source with its
high-energy power-law tail. The best fit with \emph{reflionx} that we
generated was \emph{tbabs*constant*(reflionx+diskbb),} which yields a
rX\textsuperscript{2} of 1.083; a solid chi-square and a strong model.
We froze the solar abundance ratio at 1.0, a reasonable ratio for X-ray
binaries (Arnaud, Dorman, 2015).

The reflection model photon index of 1.605 confirms our previous
power-law index: this source has a hard photon tail at high energies.
Additionally, the optical depth generated from \emph{reflionx} gives
1.05, indicating that the emission from the jets of energy coming from
the compact object is scattering off the hot particles and electrons in
the corona in high amounts, which is why we see a power law. This
reflection model, helping us understand the high energy astrophysics,
also contributes to the strong suggestion that our modeling of Swift 15
is that of a black-hole system.

\subsubsection{Light Curves and Power Spectra for Transient
\emph{Swift J174540.7-290015}}

\begin{figure}[htpb]
\centering
\includegraphics[width=1\columnwidth]{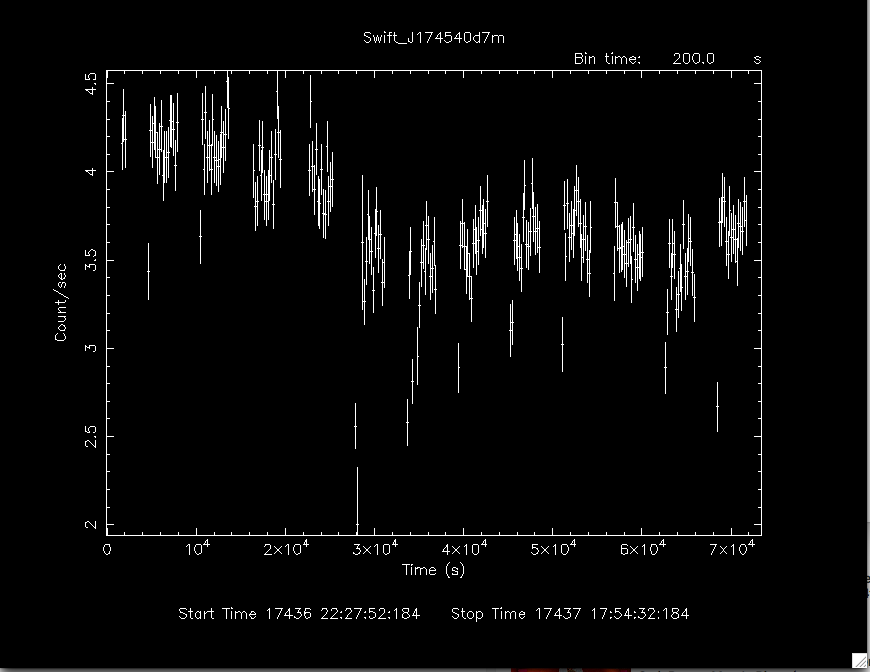}
\caption{The light curve of module A of Swift 15, with brightness
intensity (count/second) as a function of time (seconds). The bin time
is 200 seconds, with the entire duration being 24 hours. The gaps
between each cluster of data points is simply the period of time when
the rotating earth is blocking the telescope's view. }
\label{fig_sim}
\end{figure}

Observing the light curve for Swift 15, we do not see exceptionally
interesting features. There are no apparent peaks or significant
outliers in count rate throughout the duration that was observed,
indicating no type 1 X-ray bursts. The slight modulation of counts
throughout time hints at similarities to previous black hole candidate
observations, which have shown temporal variability in relatively high
count rate, whereas neutron stars tend to have very consistent count
rates in the lower count rates, with the exception of the few bursting
outliers from the type 1 X-ray bursts (Tetarenko, 2016). Overall, this
heavily leans towards the possibility of a black hole, as throughout the
seven months of Swift 15's outburst, no signs of type 1 X-ray bursts
have ever been identified.

\begin{figure}[htpb]
\centering
\includegraphics[width=1\columnwidth]{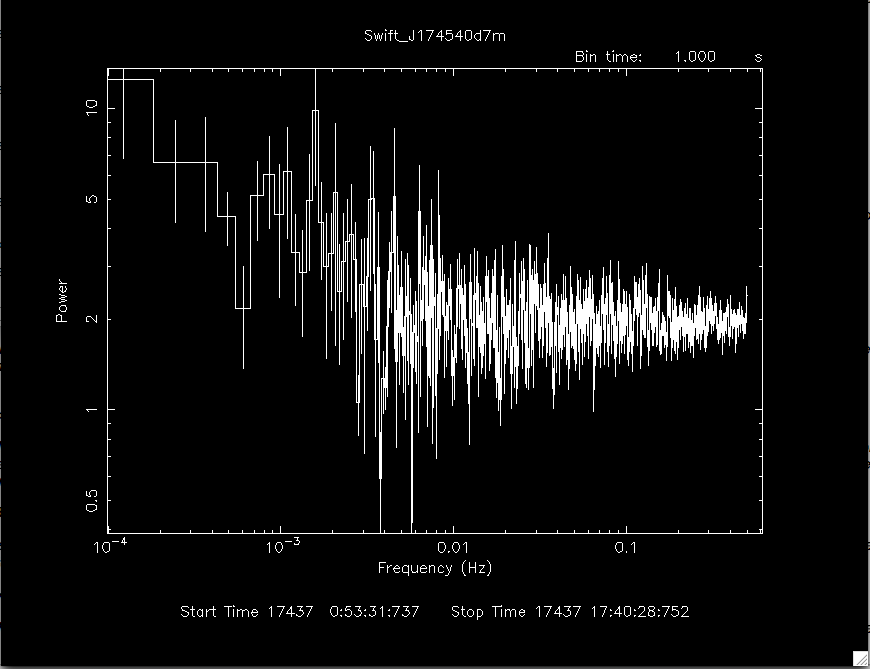}
\caption{The power spectrum of module A of Swift 15, with power
(erg/second) as a function of frequency (hertz). Rebinning of 1 second,
with maximum frequency at 0.06 Hz. Stretching the frequency range only
yields more white noise (not on this graph).}
\label{fig_sim}
\end{figure}

The power spectrum, given in Fig. 3, shows a definite red
noise accumulating from the extreme low frequencies, up to approximately
0.01 Hz, which then level off to white noise. The red noise is simply
reflective of the accretion falling from the disk onto the compact
object, and as we get into the higher frequencies, everything looks
relatively stable. Expanding the power spectrum to 500 Hz, we observe
continuous white noise. This trend is very common for most X-ray
binaries.

Some notable features in the power spectrum of Swift 15 are the slight
residual differences and small increasing peak between 0.01 Hz and 0.04
Hz, which may be indicative of a QPO at a low frequency. However, due to
the lack of professional statistics and fluctuation observations, we
cannot confirm this phenomenon. This result is still very fascinating,
as it hints and provides more suggestive evidence that Swift 15 is
likely to be a black-hole candidate.

\subsubsection{Best Fits for Transient \emph{Swift J174540.2-290037}}

\begin{figure}[htpb]
\centering
\includegraphics[width=1\columnwidth]{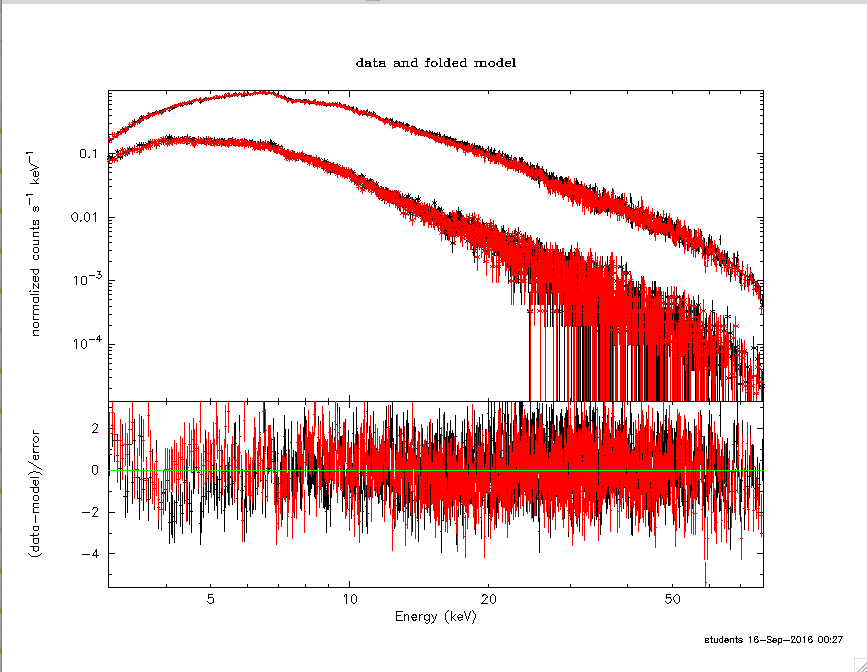}
\caption{Folded spectrum of best fit model for Swift 37, with counts as a
function of energy (keV). The datasets, upper emission (source) and
lower emission (background) are in the top region. The bottom box is the
residuals between the model and the data.}
\label{fig_sim}
\end{figure}

As shown in Fig. 4, we produced the best fit model for Swift
37, which is \emph{tbabs * constant * (powerlaw + gaussian)}. The
rX\textsuperscript{2} result is 1.091, a strong fit for the dataset.
Unlike the modeling for Swift 15, modeling for Swift 37 does not have a
diskbb component. This absence of a blackbody component is likely
because the power law photon index is 1.475; the source is at its very
hard state. At this initial hard state, we see no diskbb component in
our model, currently; however, as time progresses, it will transition
into a hard-soft state. This reasoning is logical, as Swift 37 was
detected several months after the outbursting of Swift 15.

The gaussian creates an iron line at 6.501 keV, only a slight 0.1 keV
deviation from the expected neutral-iron emission at 6.4 keV. A
noticeable feature of this gaussian component is its linewidth (Sigma)
at 0.707 keV. Similarly, to Swift 15, this source has a broad line
energy.

Using our best fit models, we calculate an equivalent width of 0.288 keV
for Swift 37; this is an incredibly small value (Parker, 2016). The
1.461 keV width we found for Swift 15 falls in the ambiguous range
between black hole and neutron star systems as previously observed;
however, Swift 37 has a very small value of 0.288 keV. This small
equivalent width, coupled with a its broad-line energy, is highly
indicative of spectral signatures of black hole LMXB systems (Tetarenko,
2016).

In the spectrum, the residuals fluctuate fairly evenly throughout the
energy scale. Again, there are no signs of cyclotron peaks at around
8-10 keV, nor at any higher energies.

Since we found such a solid result using equivalent widths and line
energies, as well the reasonable spectral residuals, there is no need to
delve deeper into the more sophisticated reflection model,
\emph{reflionx}, to analyze its photon-index parameters.

\begin{figure}[htpb]
\centering
\includegraphics[width=1\columnwidth]{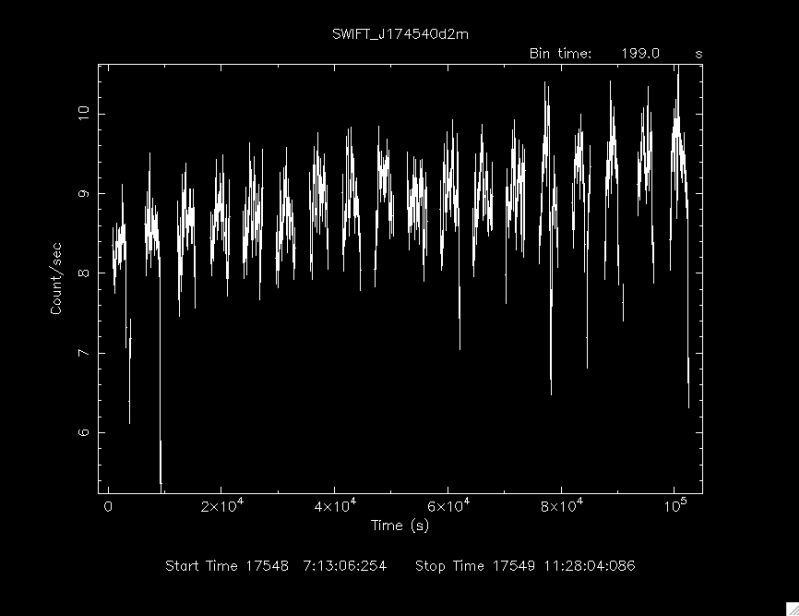}
\caption{Light curve of module A of Swift 37, with brightness intensity
(count/second) as a function of time (seconds). The bin time is 200
seconds, with the entire duration being 24 hours. The gaps in between
the cluster of datasets take account for the period of time that the
Earth is blocking the view of the telescopes into the galactic center.}
\label{fig_sim}
\end{figure}

As with the light curve for Swift 15, within this observation, Swift 37
does not show any outlier peaks indicative of type 1 X-ray bursting.
Thus, it cannot be shown to be a definitive neutron star, leading to the
strong possibility that it is a black hole system, especially that three
months worth of observations have not yielded any outbursts. Likewise
with the observation from Swift 15, Swift 37 shows trends of variability
in its light curve, indicating that this source is indeed a binary
system (which we already know). The high count rate variability,
fluctuating between 8 to 9 counts/sec, also suggests black hole
signatures, as seen with Swift 15 as well.

\begin{figure}[htpb]
\centering
\includegraphics[width=1\columnwidth]{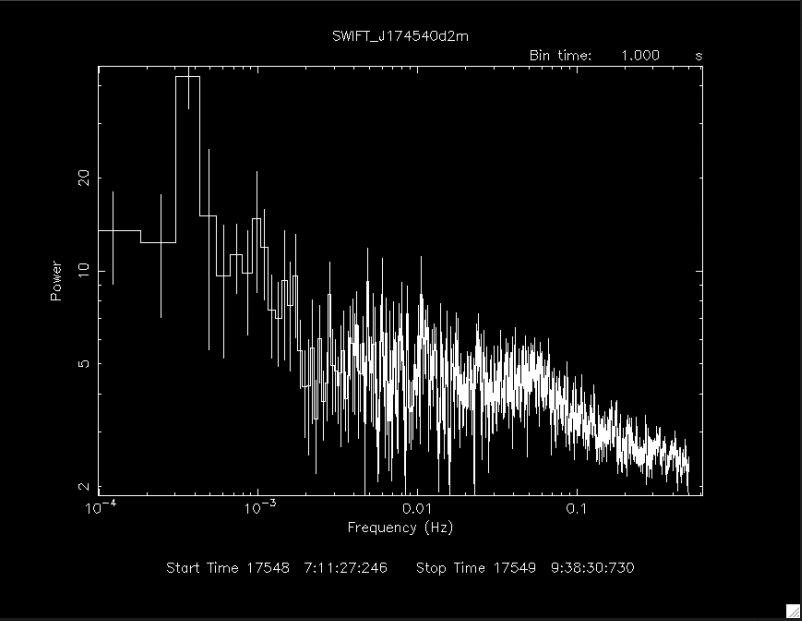}
\caption{The power spectrum of module A of Swift 37, with power
(erg/second) as a function of frequency (hertz). Rebinning of 1 second,
with maximum frequency at 0.06 Hz. Expanding the frequency range yields
solely white noise, with no other fluctuations.}
\label{fig_sim}
\end{figure}

The power spectrum for Swift 37, given in Fig. 6, shows a red
noise descending from the low frequencies up to approximately
2*10\textsuperscript{-3} Hz, which then levels off to white noise. The
white noise carries on until roughly 0.04 Hz but then shows red noise
again. Stretching this preliminary graph up to 500 Hz (not shown here),
we see a continuous and steady white noise after 1 Hz.

Zooming in to see the power spectrum up to the maximum of 0.5 Hz, we see
many interesting line features. Between 0.03 Hz and 0.05 Hz, we see a
peak in the dataset, the only noticeably significant fluctuation within
the entire power spectrum. This finding highly suggests the presence of
a low-frequency QPO, a definitive characteristic of a black hole. Again,
as with Swift 15, we need to apply astrophysical statistics, which is
beyond our research, to solidly confirm this feature.

\subsection{Conclusions and Future Work}

We have completed a thorough spectral analysis of transients \emph{Swift
J174540.7-290015} and \emph{Swift J174540.2-290037,} with model fitting,
timing analysis, and power-spectrum evaluations. Through our combined
results, we present a huge discovery: highly suggestive evidence of two
new black hole sources in the center of the galaxy.

Our modeling for transient \emph{Swift J174540.7-290015} yields a low
temperature disk blackbody spectrum coupled with a hard power-law tail,
a spectral signature that is very indicative of black-hole systems. In
addition, the model spectrum shows stably fluctuating residuals with no
presence of high-mass neutron-star cyclotron lines. The light curves did
not show any sign of type 1 X-ray bursting, but they do show some
definite count-rate variability. Also, the presence of black-hole
low-frequency quasi-periodic oscillations within the power spectrum are
definitely noticeable. The culmination of our spectral results yields
strong evidence that Swift 15 is a black-hole transient-\/-one that is
located just 17 arcseconds from the galactic center.

From our results, transient \emph{Swift J174540.2-290037} also appears
to be an extremely strong black-hole candidate. The combination of its
broad gaussian iron line and its exceptionally small equivalent width is
already a solid argument that it is a black-hole transient. Moreover,
the absence of type 1 X-ray bursts, alongside the increasing
fluctuational peak features observed in the power spectrum analysis,
possibly of a low-frequency QPO, adds further evidence that Swift 37 is
a black hole transient, only 10 arcseconds from the center of the
galaxy.

Our extremely exciting results must be pursued further by professional
astrophysicists who can solidify our identification of these two black
hole systems. Many more \emph{NuSTAR} observations need to be collected
to monitor the light curves for X-ray bursts. Furthermore, for QPO
confirmation, intense statistical analysis, that is beyond the scope of
our project, should be performed on our power spectra. Outside our
research, and to definitively categorize our transients into a source
population, dynamical mass observations must be done; however, if a
continuation of one year's observations of light curves do not show
outbursts, and the power spectra yields statistical QPO indications, our
analysis of the spectroscopy is enough to make any astrophysicist
believe that these two sources are black holes systems.

Black holes are one of the most exciting fields in X-ray astronomy due
to their enigmatic nature, and we have identified two new transients,
located extremely close to the center of the galaxy, as strong black
hole candidates! Previously, only one highly believed, unambiguous black
hole candidate has been located within two or three parsecs from the
galactic center, and now, our analysis yields the addition of two more
within this closeness to the galactic center. With less than 70 solid
black hole candidates identified throughout the entire galaxy; we now
present that three of these are within the central few parsecs of the
galaxy. This is the first highly suggestive evidence for an excess of
observed black holes clustered in the center of the galaxy!

We hope that our research will motivate other astrophysicists to further
analyze our thrilling results and to give more insight into the
longstanding challenge of deciphering the nature of black holes in our
galaxy.
 
\emergencystretch=1em

\end{document}